\documentclass[12pt,epsf]{article}
\usepackage{graphicx}
\begin{document}

\def\a{\alpha}
\def\b{\beta}
\def\c{\varepsilon}
\def\d{\delta}
\def\e{\epsilon}
\def\f{\phi}
\def\g{\gamma}
\def\h{\theta}
\def\k{\kappa}
\def\l{\lambda}
\def\m{\mu}
\def\n{\nu}
\def\p{\psi}
\def\q{\partial}
\def\r{\rho}
\def\s{\sigma}
\def\t{\tau}
\def\u{\upsilon}
\def\v{\varphi}
\def\w{\omega}
\def\x{\xi}
\def\y{\eta}
\def\z{\zeta}
\def\D{\Delta}
\def\G{\Gamma}
\def\H{\Theta}
\def\L{\Lambda}
\def\F{\Phi}
\def\P{\Psi}
\def\S{\Sigma}

\def\o{\over}
\def\beq{\begin{eqnarray}}
\def\eeq{\end{eqnarray}}
\newcommand{\gsim}{ \mathop{}_{\textstyle \sim}^{\textstyle >} }
\newcommand{\lsim}{ \mathop{}_{\textstyle \sim}^{\textstyle <} }
\newcommand{\vev}[1]{ \left\langle {#1} \right\rangle }
\newcommand{\bra}[1]{ \langle {#1} | }
\newcommand{\ket}[1]{ | {#1} \rangle }
\newcommand{\EV}{ {\rm eV} }
\newcommand{\KEV}{ {\rm keV} }
\newcommand{\MEV}{ {\rm MeV} }
\newcommand{\GEV}{ {\rm GeV} }
\newcommand{\TEV}{ {\rm TeV} }
\def\diag{\mathop{\rm diag}\nolimits}
\def\Spin{\mathop{\rm Spin}}
\def\SO{\mathop{\rm SO}}
\def\O{\mathop{\rm O}}
\def\SU{\mathop{\rm SU}}
\def\U{\mathop{\rm U}}
\def\Sp{\mathop{\rm Sp}}
\def\SL{\mathop{\rm SL}}
\def\tr{\mathop{\rm tr}}

\def\IJMP{Int.~J.~Mod.~Phys. }
\def\MPL{Mod.~Phys.~Lett. }
\def\NP{Nucl.~Phys. }
\def\PL{Phys.~Lett. }
\def\PR{Phys.~Rev. }
\def\PRL{Phys.~Rev.~Lett. }
\def\PTP{Prog.~Theor.~Phys. }
\def\ZP{Z.~Phys. }


\baselineskip 0.7cm

\begin{titlepage}

\begin{flushright}
IPMU 08-0028
\end{flushright}

\vskip 1.35cm
\begin{center}
{\large \bf
CP Violation in $B_s$ Mixing \\in the SUSY SU(5) GUT with Right-handed Neutrinos
}
\vskip 1.2cm
Junji Hisano$^{(a,b)}$ and Yasuhiro Shimizu$^{(b,c)}$
\vskip 0.4cm

{$(a)$ \em Institute for Cosmic Ray Research (ICRR), \\
University of Tokyo, Kashiwa, Chiba 277-8582, Japan}\\
{$(b)$ \em Institute for the Physics and Mathematics of the Universe (IPMU), \\
University of Tokyo, Kashiwa, Chiba 277-8568, Japan}\\
{$(c)$ \em KEK, Tsukuba, Ibaraki 305-0801,Japan }

\vskip 1.5cm

\abstract{
  It is recently announced by the {\bf Ut}{\it fit} collaboration that
  the CP phase of the $B_s$ mixing amplitude, $\phi_{B_s}$, deviates
  more than $3 \sigma$ from the standard-model prediction.  In this
  paper we discuss how large correction to $\phi_{B_s}$ is possible in
  the supersymmetric SU(5) ground unified model (SUSY SU(5) GUT) with
  right-handed neutrinos. Here, we assume the supergravity-like
  boundary condition for the SUSY-breaking terms. We found that the
  $95\%$ probability region derived by the {\bf Ut}{\it fit}
  collaboration is marginal in this model.
}
\end{center}
\end{titlepage}

\setcounter{page}{2}

In the supersymmetric standard model (SUSY SM), the SUSY-breaking
terms are introduced, and they are sources of flavor and/or CP
violation. We may get insights to physics affecting the SUSY-breaking
terms beyond the SUSY SM by studies of flavor and/or CP violating
phenomena in the hadronic and leptonic sectors.

In the supersymmetric ground unified models (SUSY GUTs), new
flavor-violating interactions are introduced, and rich
flavor-violating structure in sfermion mass matrices may be generated
by the GUT interactions. When the SUSY-breaking terms in the SUSY SM
are originated from dynamics above the GUT scale, the GUT interactions
affect them radiatively \cite{Hall:1985dx}.  In the SUSY GUTs, quarks
and leptons are embedded in common GUT multiplets. Thus, one of tests
of the SUSY GUTs is studies of correlations among flavor-violating
phenomena in the leptonic and hadronic sectors.

Recently, it is announced by the {\bf Ut}{\it fit} collaboration that the
phase of the $B_s$ mixing amplitude deviates more than $3 \sigma$ from
the SM prediction \cite{Bona:2008jn}. They parametrized
the new-physics contribution to the $B_s$ mixing as 
\begin{eqnarray}
C_{B_s}e^{2i \phi_{B_s}}&\equiv&
\frac{\langle B_s|H_{\rm eff}^{\rm full} |\bar{B_s}\rangle}
{\langle B_s|H_{\rm eff}^{\rm SM} |\bar{B_s}\rangle} \ ,
\end{eqnarray}
and performed the model-independent analysis using the available
observables of the $B_s$ system. They found that the $68\%~(95\%)$
probability regions of $\phi_{B_s}$ and $C_{B_s}$ are given as 
\begin{eqnarray}
\phi_{B_s} &=&\left\{
\begin{tabular}{l}
$[-25.5^{\circ},~-14.3^{\circ}]~([-30.45^{\circ},~-9.29^{\circ}])$\\
$[-73.1^{\circ},~-63.3^{\circ}]~([-78.45^{\circ},~-58.2^{\circ}])$
\end{tabular}
\ , \right. 
\nonumber\\
C_{B_s}&=&[0.78,~1.36]~([0.62,~1.93]) \ .
\label{utfit}
\end{eqnarray}
The phase $\phi_{B_s}$ deviates from zero at $3.7\sigma$.  The further
studies are necessary so that the deviation is confirmed, since their
result comes from a combination of various observables. 
They also used the SU(3) symmetry in order to evaluate unknown strong
phases.  However, their result encourages us to reanalyze the
new-physics contribution to various flavor-changing neutral current
(FCNC) processes.

In this paper we discuss how large deviation of the phase $\phi_{B_s}$
is possible in the SUSY SU(5) GUT with right-handed neutrinos. Here,
we assume the supergravity-like (or CMSSM-like) boundary condition for
the SUSY-breaking terms.  The large mixing angle observed in the
atmospheric neutrino experiments suggests that the sizable deviations
from the SM may appear in the bottom-strange quark transitions in the
SUSY GUTs. The neutrino Yukawa interaction radiatively generates the
bottom-strange component in the right-handed down-squark mass matrix
\cite{Moroi:2000tk}. On the other hand, the contribution to the
bottom-strange quark transition is constrained from the upperbound on
$Br(\tau\rightarrow \mu \gamma)$ in the model \cite{Hisano:2003bd}.
The neutrino Yukawa interaction also induces the lepton-flavor
violating terms in the left-handed slepton mass matrix so that
lepton-flavor violating processes are predicted
\cite{Borzumati:1986qx}, and $Br(\tau\rightarrow \mu \gamma)$ is also
enhanced by the large atmospheric neutrino-mixing angle
\cite{Hisano:1998fj}. We evaluate $\phi_{B_s}$ in the model, and it is
found that it can deviate at most $\sim(8^\circ-9^\circ)$ from zero
under the experimental constraint.  Thus, it is difficult to get such
a large deviation in $\phi_{B_s}$ from the SM as in the $68\%$
probability regions derived by the {\bf Ut}{\it fit} collaboration in
Eq.~(\ref{utfit}), while one of the $95\%$ probability region is
marginal.

The SUSY contribution to the phase in the $B_s$ mixing is also
evaluated by
Refs.~\cite{Ciuchini:2003rg,Parry:2007fe,Cheung:2007pj,Goto:2007ee} in
a framework of the SUSY GUTs. In Ref.~\cite{Ciuchini:2003rg} the
constraints on the mass insertion parameters for the right-handed
squarks are derived from the lepton-flavor violating processes using
the GUT relation among the squark and slepton mass matrices at the GUT
scale.  The authors in Refs~\cite{Parry:2007fe,Cheung:2007pj} evaluate
the phase in the $B_s$ mixing in the SUSY GUTs, though they do not
include the Higgs boson mass bound in their analysis. The analysis in
Ref.~\cite{Goto:2007ee} is similar to ours, though they do not
evaluate the maximum value of the phase in the $B_s$ mixing in the
model.

First, we briefly explain the SUSY SU(5) GUT with right-handed
neutrinos. In the model, quarks and leptons in the SUSY SM are {\bf
  10}- and ${\bf 5^\star}$-dimensional multiplets, while the
right-handed neutrinos are singlets.  The Yukawa couplings for quarks
and leptons and the Majorana mass terms for the right-handed neutrinos
in this model are given as
\begin{eqnarray}
W&=& 
\frac14 f_{ij}^{u} \Psi_i \Psi_j H 
+\sqrt{2} f_{ij}^{d} \Psi_i \Phi_j \overline{H}
+f_{ij}^{\nu} \Phi_i \overline{N}_j {H}
+M_{ij} \overline{N}_i \overline{N}_j,
\label{superp_gut}
\end{eqnarray}
where $\Psi$ and $\Phi$ are the {\bf 10}- and {$\bf 5^\star$}-dimensional
multiplets, respectively, and $\overline{N}$ is the right-handed
neutrinos.  $H$ ($\overline{H}$) is {\bf 5} ({$\bf 5^\star$})-
dimensional Higgs multiplets. After removing the unphysical degrees of
freedom, the Yukawa coupling constants in Eq.~(\ref{superp_gut}) are
given as follows,
\begin{equation}
f^u_{ij} =
V_{ki} f_{u_k} {\rm e}^{i \varphi_{u_k}}V_{kj}, \ ~~
f^d_{ij} = f_{d_i} \delta_{ij}, \ ~~
f^\nu_{ij} = {\rm e}^{i \varphi_{d_i}} 
U^\star_{ij} f_{\nu_j} ,
\end{equation}
where $\sum_i \varphi_{f_i} =0$ $(f=u,d)$.  The unitary matrix $V$
is the CKM matrix in the extension of the SM to the SUSY SU(5)
GUT. When the Majorana mass matrix for the right-handed neutrinos $M$
is diagonal in the above basis, $U$ is the the MNS matrix (with the
Majorana phases), since the left-handed neutrino mass matrix is
\begin{eqnarray}
(m_\nu)_{ij}
&=&
(f^{\nu}(M^{-1}) f^{\nu T})_{ij} \langle H_f \rangle^2 \ .
\end{eqnarray}
Here, $H_f$ is a doublet Higgs in $H$. In the following, we assume for
simplicity that $M$ is diagonal, $M={\rm
  diag}(M_{N_1},M_{N,2},M_{N_3})$.

The colored-Higgs multiplets $H_c$ and $\overline{H}_c$ are introduced
in $H$ and $\overline{H}$ as SU(5) partners of the Higgs doublets in
the SUSY SM, $H_f$ and $\overline{H}_f$, respectively, and they have new
flavor-violating interactions given by Eq.~(\ref{superp_gut}).  When the
SUSY-breaking terms are generated above the GUT scale, the
colored-Higgs interactions give corrections to the sfermion mass
matrices. When the supergravity-like boundary condition is assumed,
the SUSY-breaking scalar masses and the trilinear SUSY-breaking terms
at the reduced Planck scale ($M_G$) are given universally by $m_0$ and
$A_0$, respectively. In this case, the off-diagonal terms ($i\ne j$)
in the sfermion mass matrices at low energy are approximated as
\begin{eqnarray}
(m_{{Q}}^2)_{ij}  &\simeq&-\frac{2}{(4\pi)^2} 
V_{ki}^\star
f_{u_k}^2
V_{kj} (3m_0^2+ A_0^2) (3 \log\frac{M_G}{M_{GUT}}
                        + \log\frac{M_{GUT}}{M_{SUSY}}),\nonumber\\
(m_{\overline{U}}^2)_{ij}  &\simeq&-\frac{4}{(4\pi)^2} 
{\rm e}^{i\varphi_{u_i}}V_{ik}
f_{d_k}^2
V_{jk}^\star {\rm e}^{-i\varphi_{u_j}} (3m_0^2+ A_0^2)
\log\frac{M_G}{M_{GUT}}, \nonumber\\
(m_{\overline{D}}^2)_{ij}  &\simeq&-\frac{2}{(4\pi)^2} 
{\rm e}^{-i\varphi_{d_i}}U_{ik}
f_{\nu_k}^2
U^\star_{jk} {\rm e}^{i\varphi_{d_j}} (3m_0^2+A_0^2) 
\log\frac{M_G}{M_{GUT}},\nonumber\\
(m_{L}^2)_{ij}  &\simeq&-\frac{2}{(4\pi)^2} 
U_{ik}
f^2_{\nu_k} 
U_{jk}^\star(3m_0^2+ A_0^2) 
\log\frac{M_G}{M_{N_k}},\nonumber\\
(m_{\overline{E}}^2)_{ij}  &\simeq&-\frac{6}{(4\pi)^2} 
{\rm e}^{-i\varphi_{d_i}} V^\star_{ki}
f_{u_k}^2
V_{kj} {\rm e}^{i\varphi_{d_j}} (3m_0^2+ A_0^2)
\log\frac{M_G}{M_{GUT}},
\label{sfermionmass}
\end{eqnarray}
where $(m_{{Q}}^2)$ ($(m_{{\overline{U}}}^2)$,
$(m_{{\overline{D}}}^2)$) is left-handed (right-handed) squark mass
matrix(es) and $(m_{{L}}^2)$ ($(m_{{\overline{E}}}^2)$) is for
left-handed (right-handed) sleptons. Here, $M_{GUT}$ and $M_{SUSY}$
are the GUT scale and the SUSY-breaking scale in the SUSY SM,
respectively.  While the off-diagonal terms in the left-handed squark
and slepton mass matrices are generated in the SUSY SM with the
right-handed neutrinos, those for right-handed squarks and
sleptons are generated by the colored-Higgs interactions.

As mentioned above, the bottom-strange quark and $\tau$-$\mu$
transitions are correlated in the SUSY SU(5) GUT with right-handed
neutrinos. This is because a following simple relation is valid from
Eq.~(\ref{sfermionmass}), 
\begin{eqnarray}
\frac{(m_{\overline{D}}^2)_{23}}{(m_{{L}}^2)_{23}} 
&\simeq&
{\rm e}^{-i(\varphi_{d_2}-\varphi_{d_3})}
\frac{\log\frac{M_G}{M_{GUT}}}{\log\frac{M_G}{M_{N_3}}} \ ,
\end{eqnarray}
under an assumption, $f_{\nu_\tau}^2\gg f_{\nu_\mu}^2\gg f_{\nu_e}^2$.
The corrections to this relation are possible, when the the
right-handed neutrino mass matrix is not diagonal.  However, it is
shown in Ref.~\cite{Hisano:2003bd} that this relation is approximately
valid when the right-handed neutrino masses are hierarchical. The
hierarchical spectrum is welcome from a phenomenological viewpoint,
since the muon-neutrino Yukawa coupling is so small that
$Br(\mu\rightarrow e \gamma)$ is suppressed below the experimental
bound. Here, we also assume $U_{e3}\lsim 0.01$ in order to 
suppress $Br(\mu\rightarrow e \gamma)$.
In such a case the large corrections to
$(m_{\overline{D}}^2)_{23}$ and $(m_{{L}}^2)_{23}$ by the
colored-Higgs interactions may be allowed.

Now we evaluate how large deviation of the phase $\phi_{B_s}$ in the
$B_s$ mixing is possible in the SUSY SU(5) GUT with right-handed
neutrinos. The input parameters for the SUSY breaking parameters are
$m_0$, $A_0$, $M_{1/2}$ and $\tan\beta$. Here, $M_{1/2}$ is the
gaugino mass at $M_G$. We scan following parameter regions,
\begin{eqnarray}
0<m_0,~ M_{1/2}<3~{\rm TeV},\nonumber\\
A_0/m_0=-2,~0,~2,\nonumber\\
\tan\beta=5,~10,~30\ .
\label{region}
\end{eqnarray}
For the neutrino sector, we take $m_{\nu_\tau}=5\times 10^{-2}$~eV,
$U_{23}=1/\sqrt{2}$, and $M_{N_3}=6\times 10^{14}$~GeV.  We evaluate
the SUSY mass spectrum and the interactions by numerically solving the
renormalization-group equations for the SUSY breaking parameters in
the SUSY GUT and the SUSY SM (with right-handed neutrinos). 

In the following, we give constraints on the above parameter space
from the experimental bounds on the Higgs boson mass $m_h$ in addition
to $Br(\tau\rightarrow \mu\gamma)$.  The SUSY contribution to the
$B_s$ mixing is dominated by the box diagrams including gluino
exchanges when both left- and right-handed down-type squarks have
flavor-violating mass terms \cite{Ball:2003se}.  Thus, the SUSY
correction is less sensitive to $\tan\beta$. On the other hand,
$Br(\tau\rightarrow \mu \gamma)$ is proportional to $\tan^2\beta$
since it is generated by the effective dipole operators. The
constraint on $(m_{\overline{D}}^2)_{23}$ is more severe for larger
$\tan\beta$ due to the null results in the searches for
$\tau\rightarrow\mu\gamma$. The larger correction to the $B_s$ mixing
is possible when $\tan\beta$ is not large. The size is, however,
limited by the Higgs boson mass bound in the case.  The Higgs boson
mass bound pushes up sfermion masses including top squarks, especially
for small $\tan\beta$.

\begin{figure}[t]
\begin{center}
\begin{tabular}{cc}
\includegraphics[scale=0.35, angle = 0]{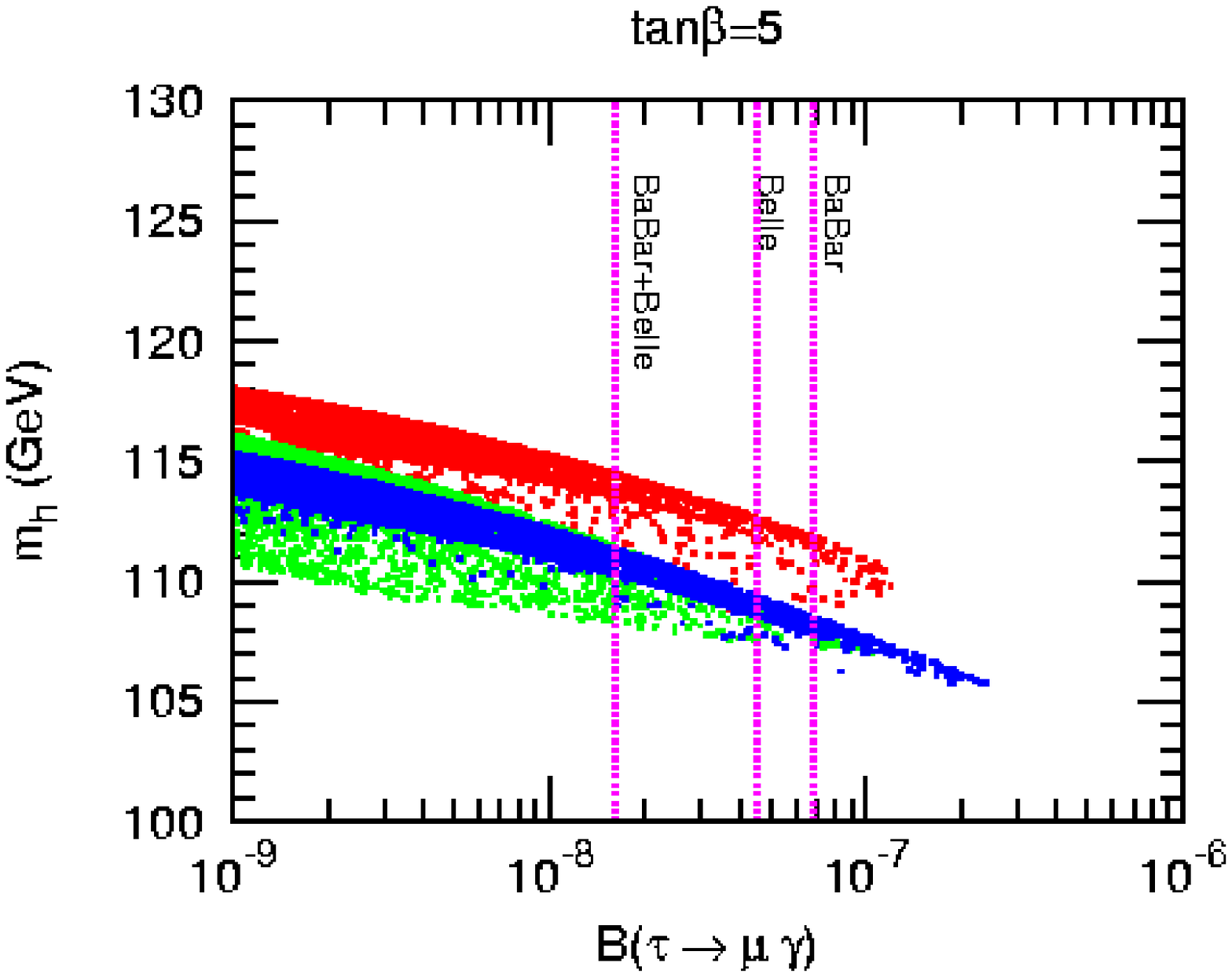}&
\includegraphics[scale=0.35, angle = 0]{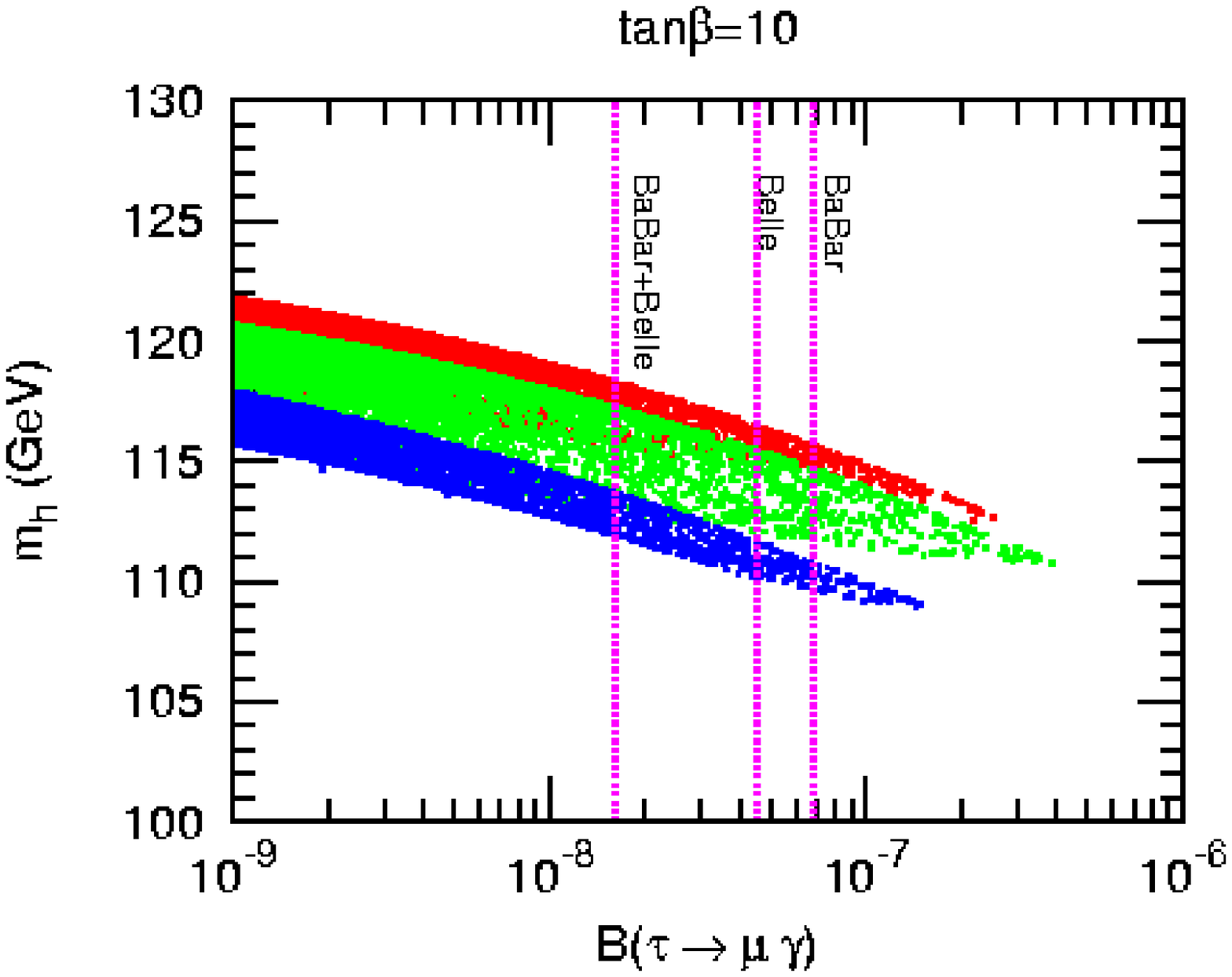}
\end{tabular}\\
\includegraphics[scale=0.35, angle = 0]{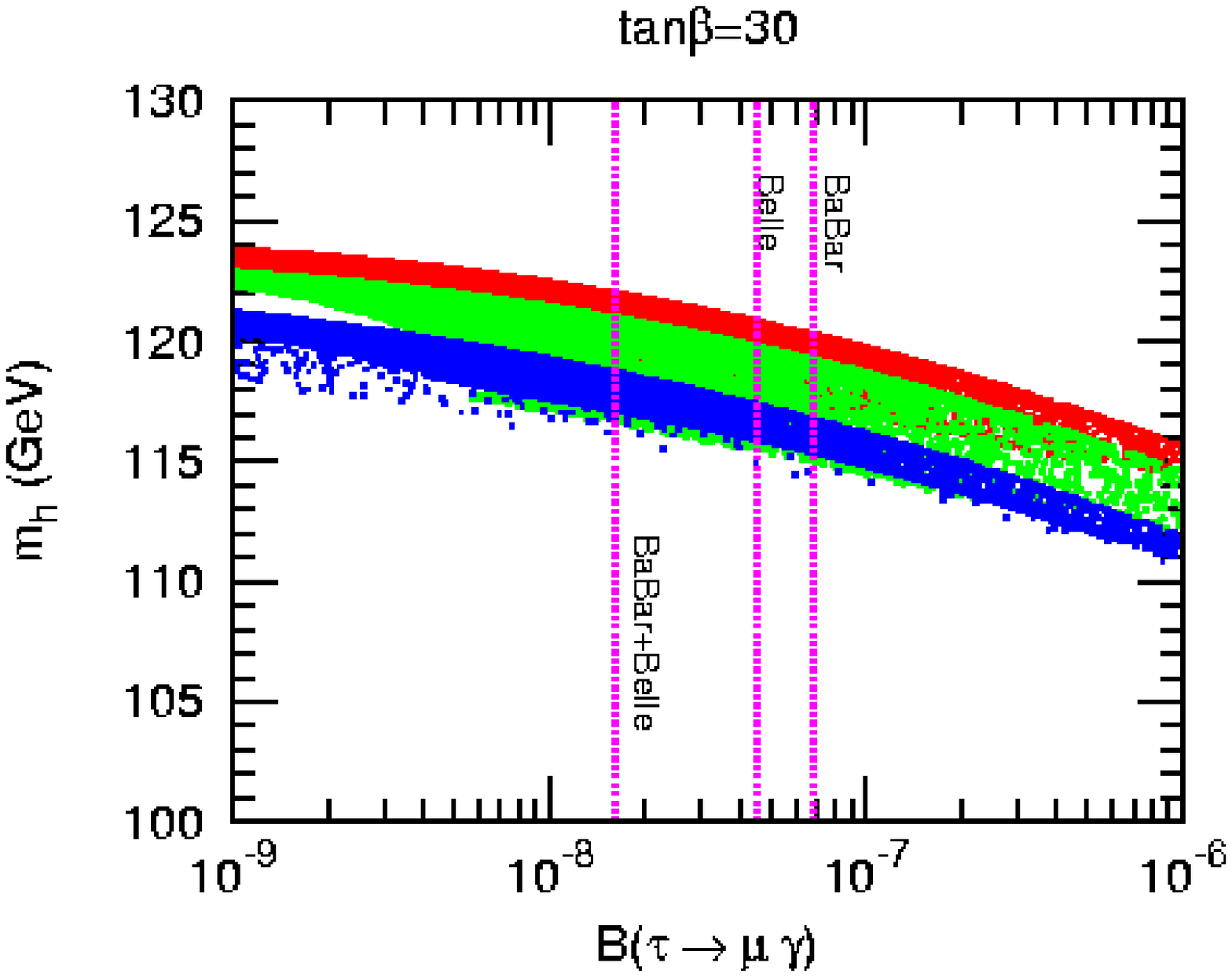}
\caption{\label{fig:tmg-mh} 
Scatter plots on plane of
$Br(\tau\rightarrow \mu\gamma)$ and $m_h$ for $\tan\beta=5,~10,~30$.
Red, green, and blue points are for $A_0/m_0=-2,~0,~+2$. 
Vertical lines are the upperbounds on $Br(\tau\rightarrow \mu\gamma)$
derived in BaBar and  Belle, and the combined one.
}
\end{center}
\end{figure}

First, we show scatter plots on a plane of $Br(\tau\rightarrow
\mu\gamma)$ and $m_h$ in Fig.~\ref{fig:tmg-mh}, in order to show that
the large and small $\tan\beta$ regions are constrained from
$Br(\tau\rightarrow\mu\gamma)$ and $m_h$, respectively, in the SUSY
SU(5) GUT with large neutrino Yukawa interaction. The current
upperbound on $Br(\tau\rightarrow \mu\gamma)$ is derived as $6.8\times
10^{-8}$ \cite{Aubert:2005ye} and $4.5\times 10^{-8}$ by BaBar and
Belle, respectively, \cite{Abe:2006sf}.  The combined upperbound of them
is $1.6\times 10^{-8}$ \cite{Banerjee:2007rj}.  The SM Higgs boson
mass lowerbound is 114.4~GeV \cite{Schael:2006cr}. In the following we
impose $m_h>111.4$~GeV \cite{Heinemeyer:2004gx}, taking theoretical
uncertainties in the Higgs boson mass evaluation. The constraint from
$Br(b\rightarrow s\gamma)$ is also included in our analysis, though we
found that it is not significant compared with
$Br(\tau\rightarrow\mu\gamma)$.

It is found from Fig.~\ref{fig:tmg-mh} that the prediction
for $Br(\tau\rightarrow \mu\gamma)$ is larger and the model is
strongly constrained by the experimental upperbound when $\tan\beta$
is larger.  On the other hand, when $\tan\beta=5$, the Higgs boson
mass bound also gives a severe constraint on the model. When $A_0$ has a
negative value, the $A_t$ term in the stop mass matrix is sizable so
that the Higgs boson becomes heavier. Then, larger value of $A_0$ with
negative sign is favored from the Higgs boson mass bound.

\begin{figure}[t]
\begin{center}
\begin{tabular}{cc}
\includegraphics[scale=0.35, angle = 0]{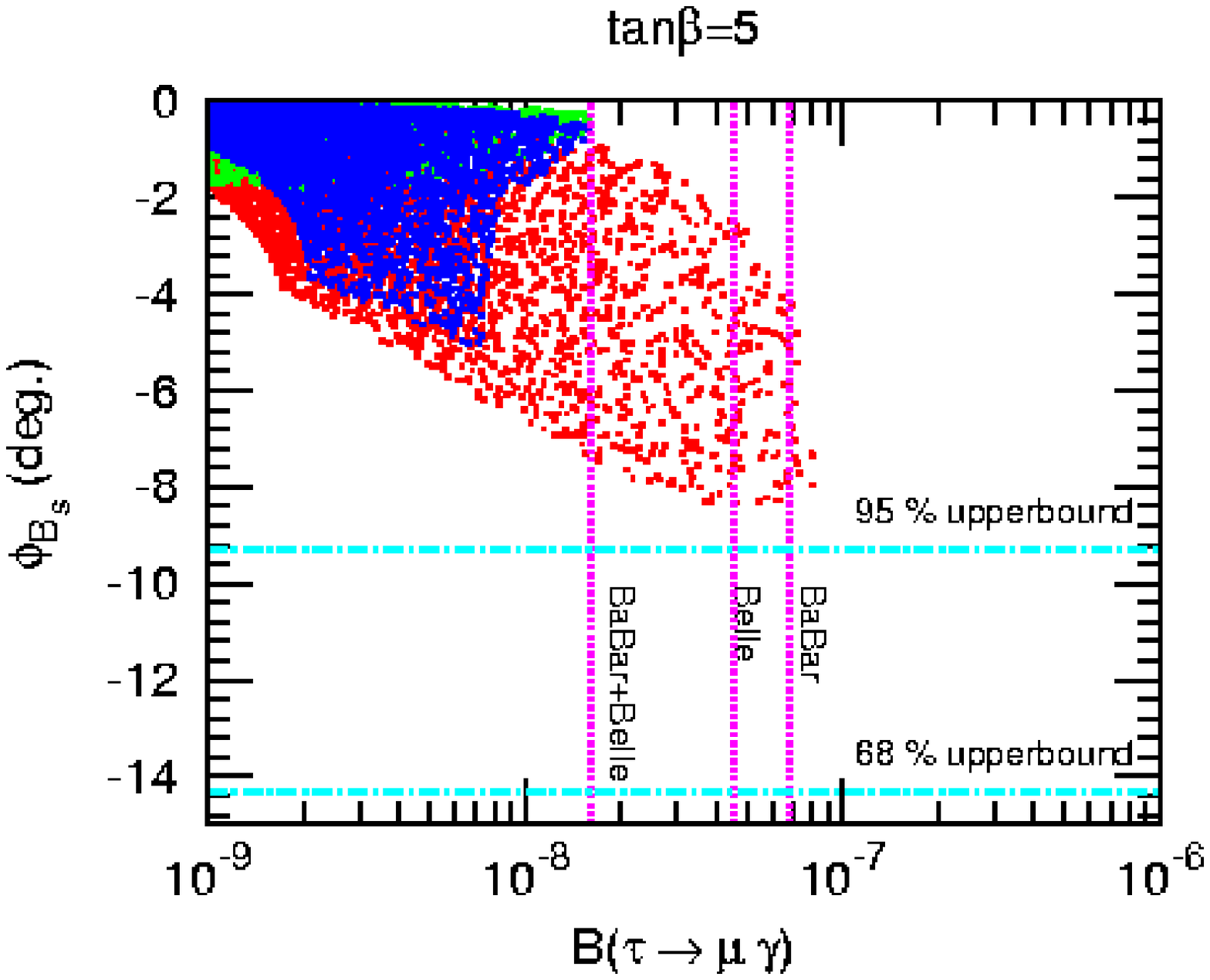}&
\includegraphics[scale=0.35, angle = 0]{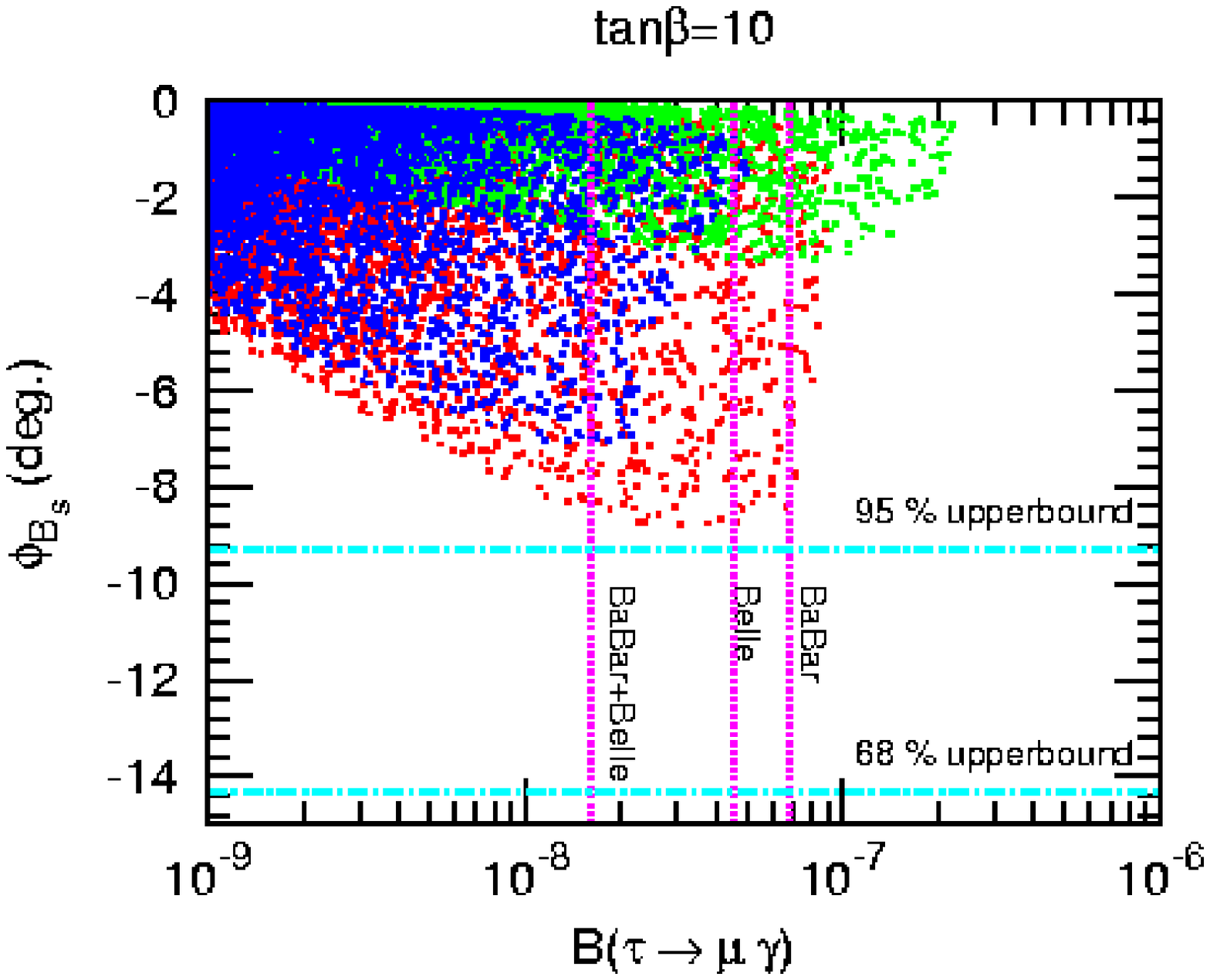}
\end{tabular}\\
\includegraphics[scale=0.35, angle = 0]{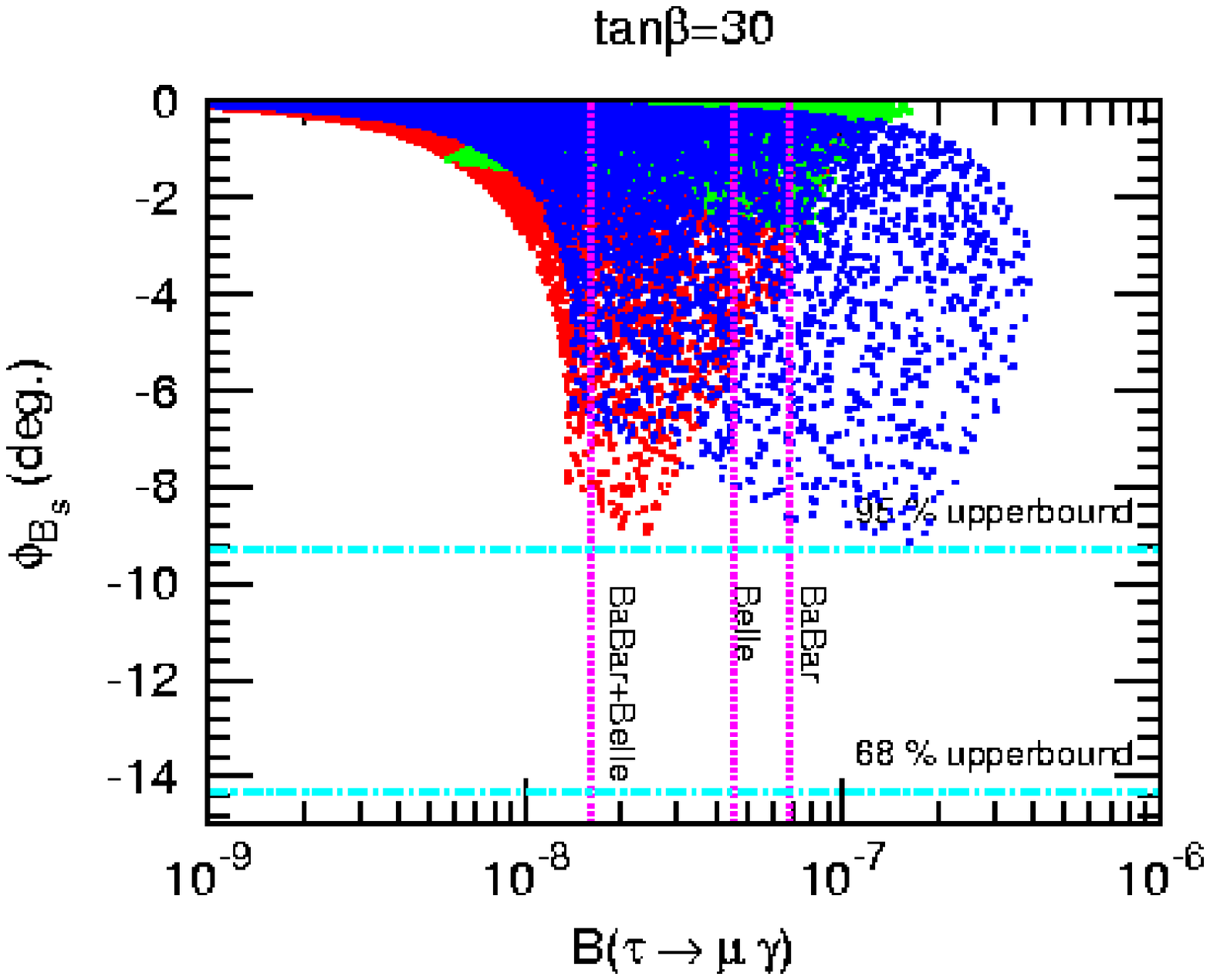}
\caption{\label{fig:tmg-fbs} 
Scatter plots on plane of $\phi_{B_s}$
and $Br(\tau\rightarrow\mu\gamma)$. Here, we imposed
the Higgs boson mass bound. Others are the same as in Fig.~\ref{fig:tmg-mh}. 
Horizontal lines show $68\%$ and $95\%$ probability regions of $\phi_{B_s}$
derived by the {\bf Ut}{\it fit} collaboration.
}
\end{center}
\end{figure}

Now we show the correlation between $\phi_{B_s}$ and
$Br(\tau\rightarrow\mu\gamma)$ in Fig.~\ref{fig:tmg-fbs}.  In the
figure we imposed the Higgs mass constraint. The deviation of
$\phi_{B_s}$ from zero is limited by $Br(\tau\rightarrow\mu\gamma)$,
especially when $\tan\beta=30$. When $\tan\beta=5$, the constraint
from $Br(\tau\rightarrow \mu\gamma)$ is not strong, however, the size of
$\phi_{B_s}$ is limited.  While heavier SUSY particle mass spectrum
gives heavier Higgs boson mass, it suppresses the SUSY contribution to
the FCNC processes. The largest deviation of $\phi_{B_s}$ from zero is
$\phi_{B_s}\simeq -(8^\circ-9^\circ)$ when $\tan\beta =10$.  Thus, it
is found that one of the $95\%$ probability region derived by the {\bf
  Ut}{\it fit} collaboration in Eq.~(\ref{utfit}) is marginal.

\begin{figure}[t]
\begin{center}
\begin{tabular}{cc}
\includegraphics[scale=0.35, angle = 0]{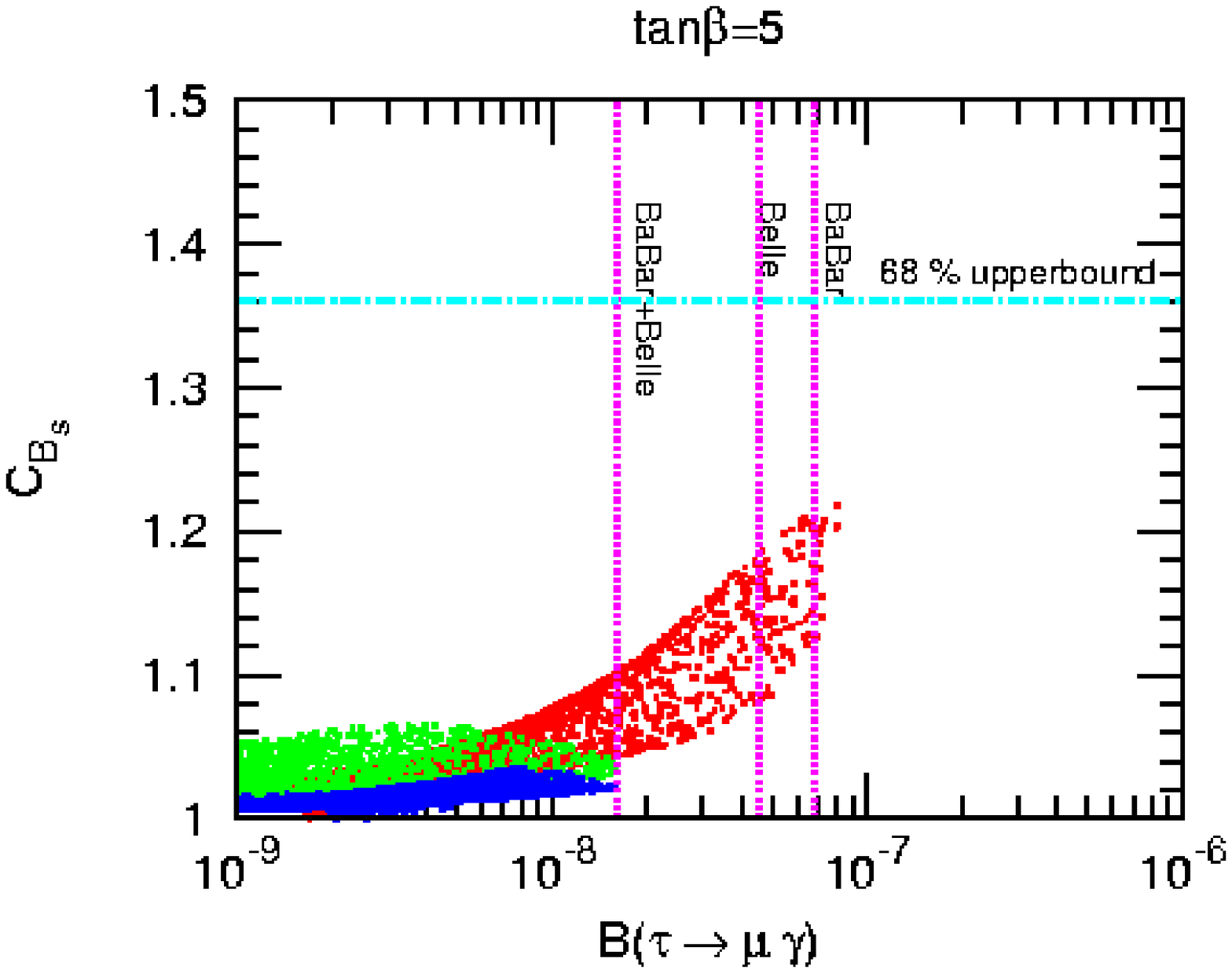}&
\includegraphics[scale=0.35, angle = 0]{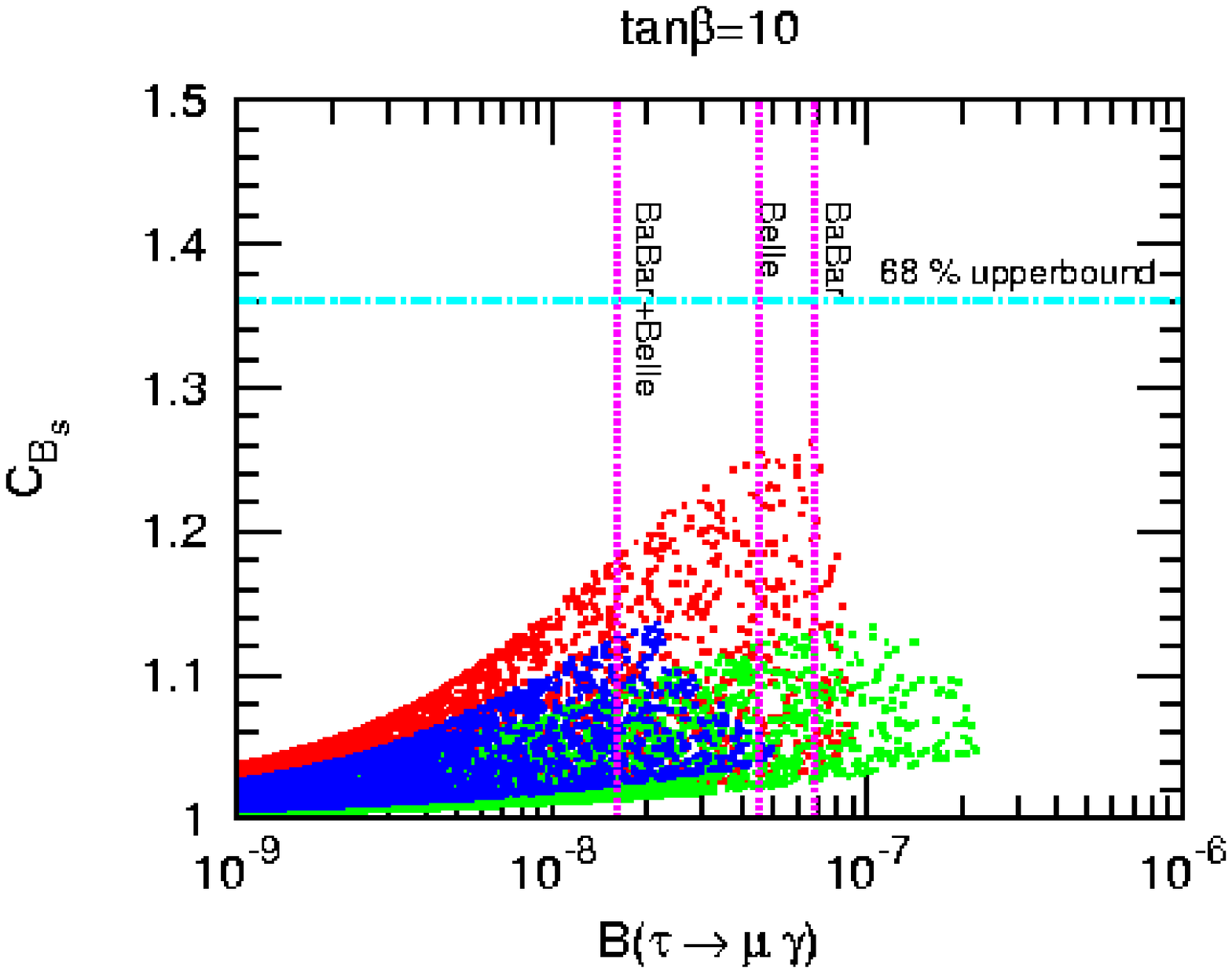}\\
\includegraphics[scale=0.35, angle = 0]{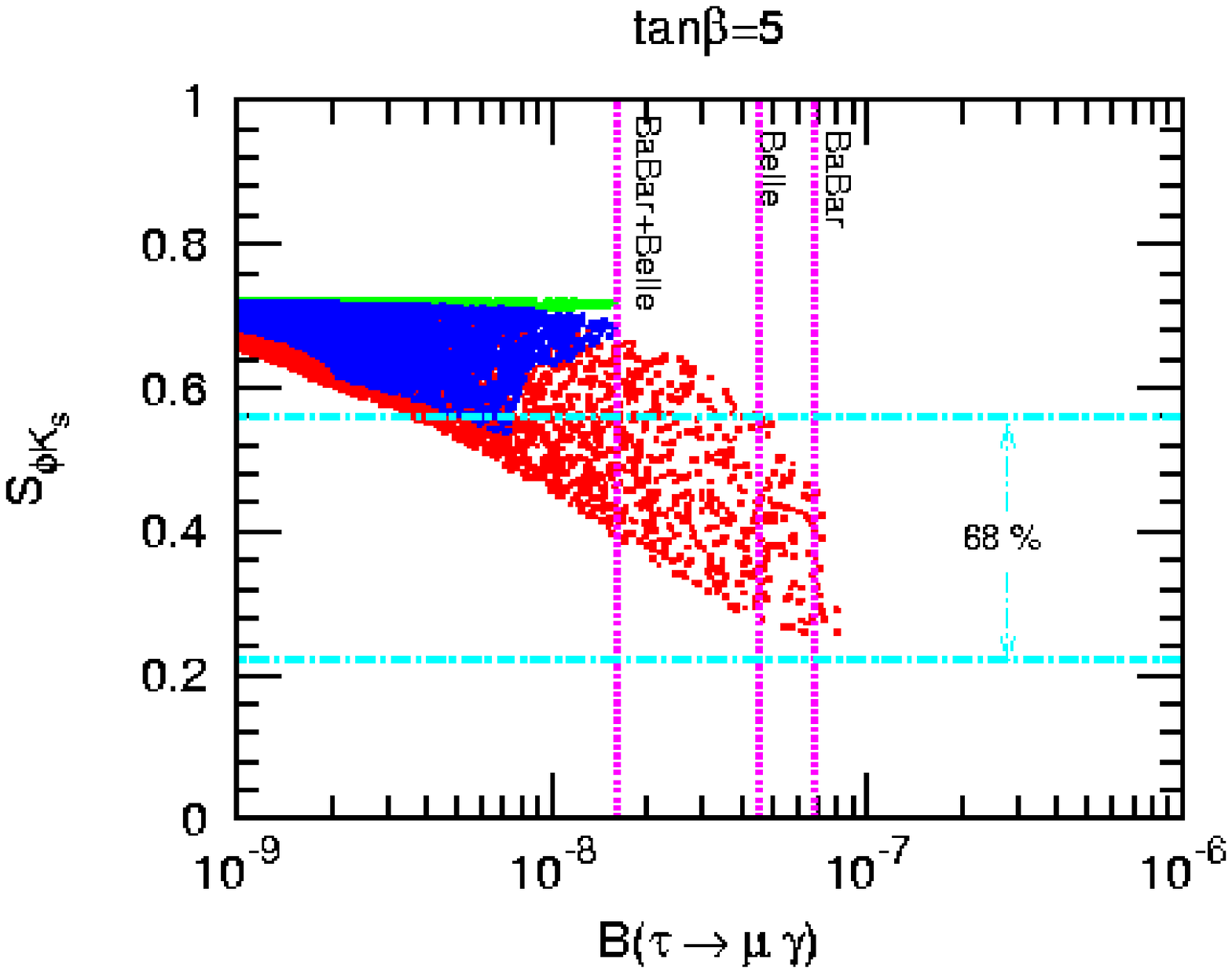}&
\includegraphics[scale=0.35, angle = 0]{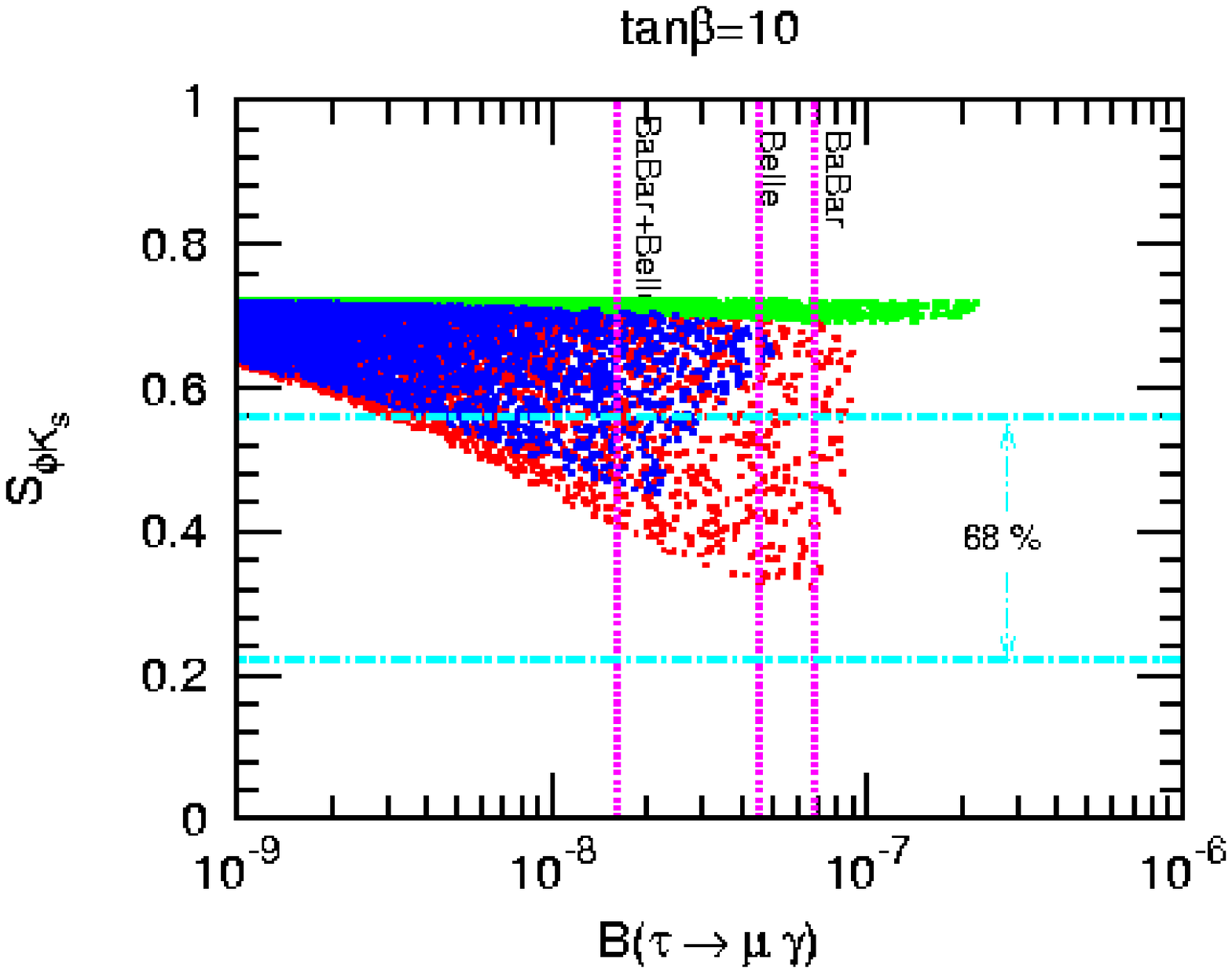}\\
\includegraphics[scale=0.35, angle = 0]{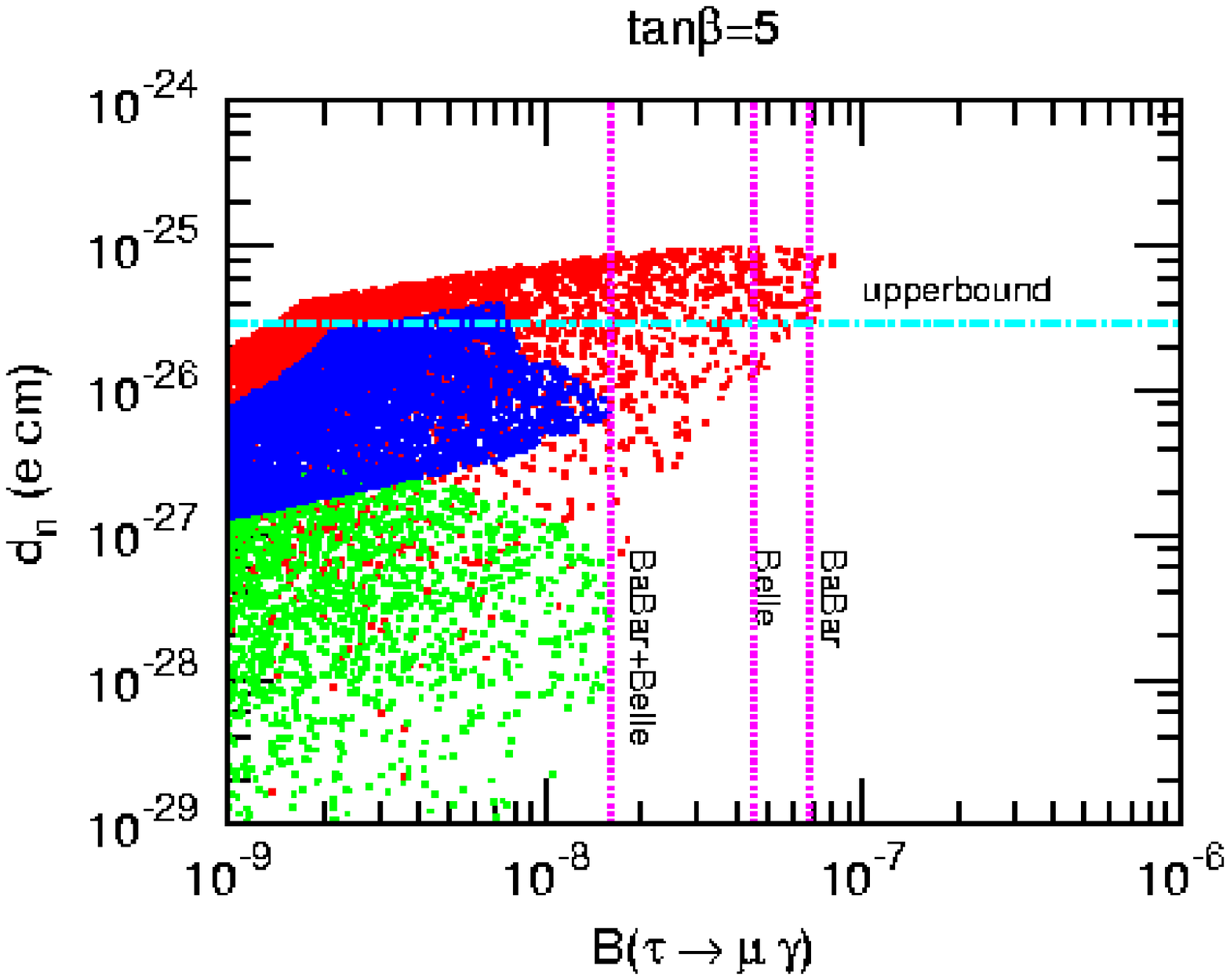}&
\includegraphics[scale=0.35, angle = 0]{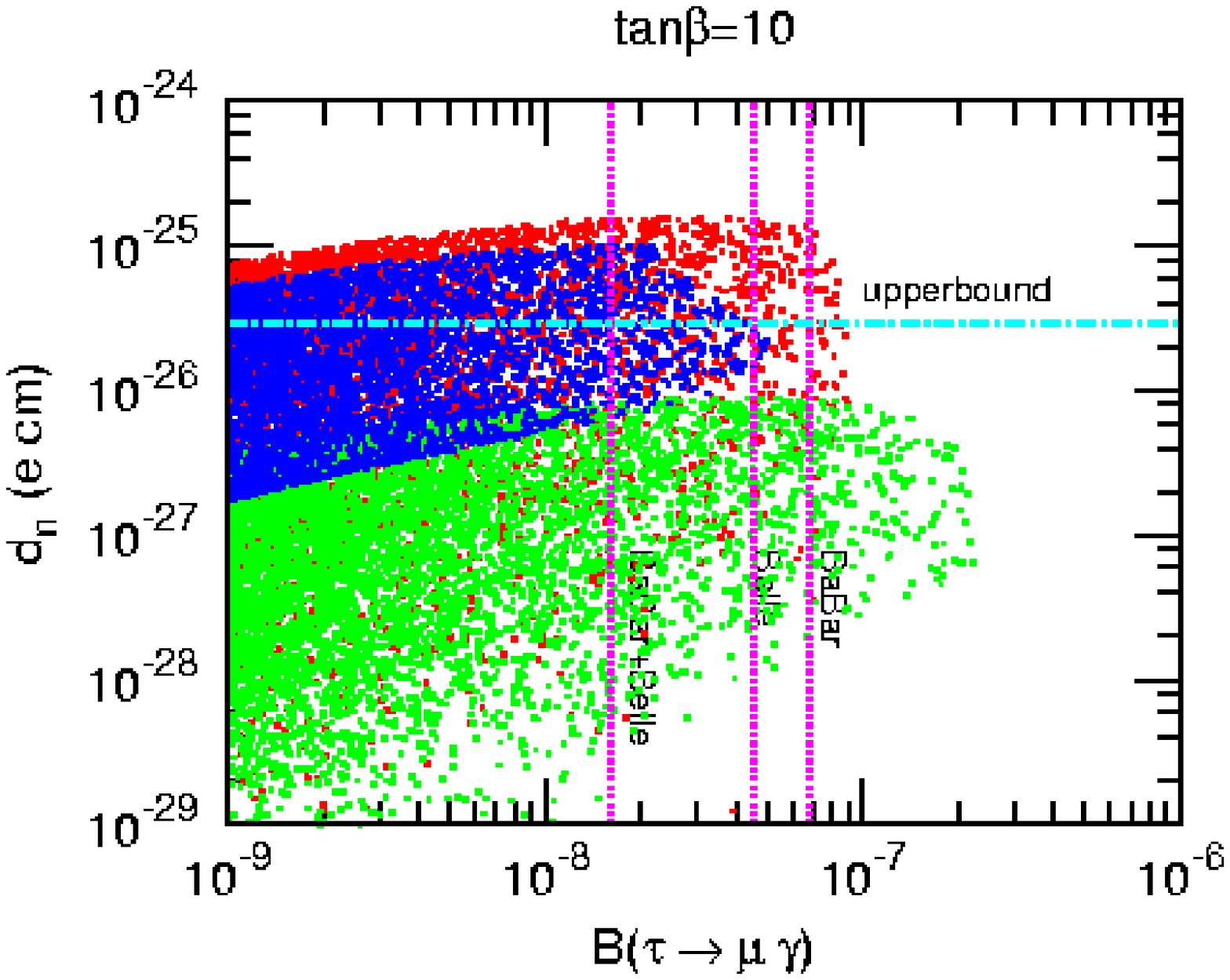}
\end{tabular}
\caption{\label{fig:tmg-spk} 
The mass difference of the $B_s$ mesons normalized by the SM prediction 
($C_{B_s}$), the CP asymmetry in $B_d\rightarrow \phi K_s$ ($S_{\phi K_s}$), 
and the neutron EDM induced by the strange CEDM ($d_n$)
are shown for $\tan\beta=5$ and 10. The experimental bounds on them
are shown by the horizontal lines in the figures.  Others are the same as in Fig.~\ref{fig:tmg-fbs}.
}
\end{center}
\end{figure}

Now we check the consistency with other observables. First, the mass
difference of the $B_s$ mesons normalized by the SM prediction,
$C_{B_s}(\equiv \Delta M_s/\Delta M_s|_{\rm SM})$, is shown in
Fig.~\ref{fig:tmg-spk}. Here, we take $\tan\beta=5$ and 10.  While
$\Delta M_s$ is precisely measured by CDF \cite{Abulencia:2006ze}, the
hadronic uncertainties in the theoretical prediction are large. The
constraint on $C_{B_s}$ is given in Eq.~(\ref{utfit}). It is found
that this constraint is not significant in
Fig.~\ref{fig:tmg-spk}. Next, the CP asymmetry in $B_d\rightarrow \phi
K_s$, which is induced by the $b$-$s$ penguin diagrams, is also shown
in Fig.~\ref{fig:tmg-spk}. The correction to the process can be as
large as $\sim 30\%$. The experimental and theoretical uncertainties
\cite{bspenguin,bspenguinth} in the $b$-$s$ penguin processes are
still large, so that we could not give a rigid constraint on the
parameters from it.

When the flavor-violating mass terms for the right-handed squarks are
non-vanishing, the hadronic EDMs are generated at one-loop
\cite{Dimopoulos:1994gj} and two-loop levels \cite{Hisano:2006mj}.  The
non-zero $(m_{\overline{D}}^2)_{23}$ generates the strange-quark
chromoelectric dipole moment, which contributes to the hadronic EDMs
\cite{Hisano:2003iw,Hisano:2004tf}.  When evaluating the neutron EDM
using the formula in Ref.~\cite{Hisano:2004tf}, it can be as large as
$\sim 10^{-25}e$~cm even in the region allowed by the
$Br(\tau\rightarrow \mu \gamma)$ and $m_h$.
(See Fig.~\ref{fig:tmg-spk}). The current experimental bound is $2.9\times
10^{-26}e$~cm \cite{Baker:2006ts}. Thus, the sizable deviation of
$\phi_{B_s}$ from the SM prediction is possible if the neutron EDM
suffers from the hadronic uncertainties or it is suppressed due to an
accidental cancellation.

\begin{figure}[t]
\begin{center}
\includegraphics[scale=0.35, angle = 0]{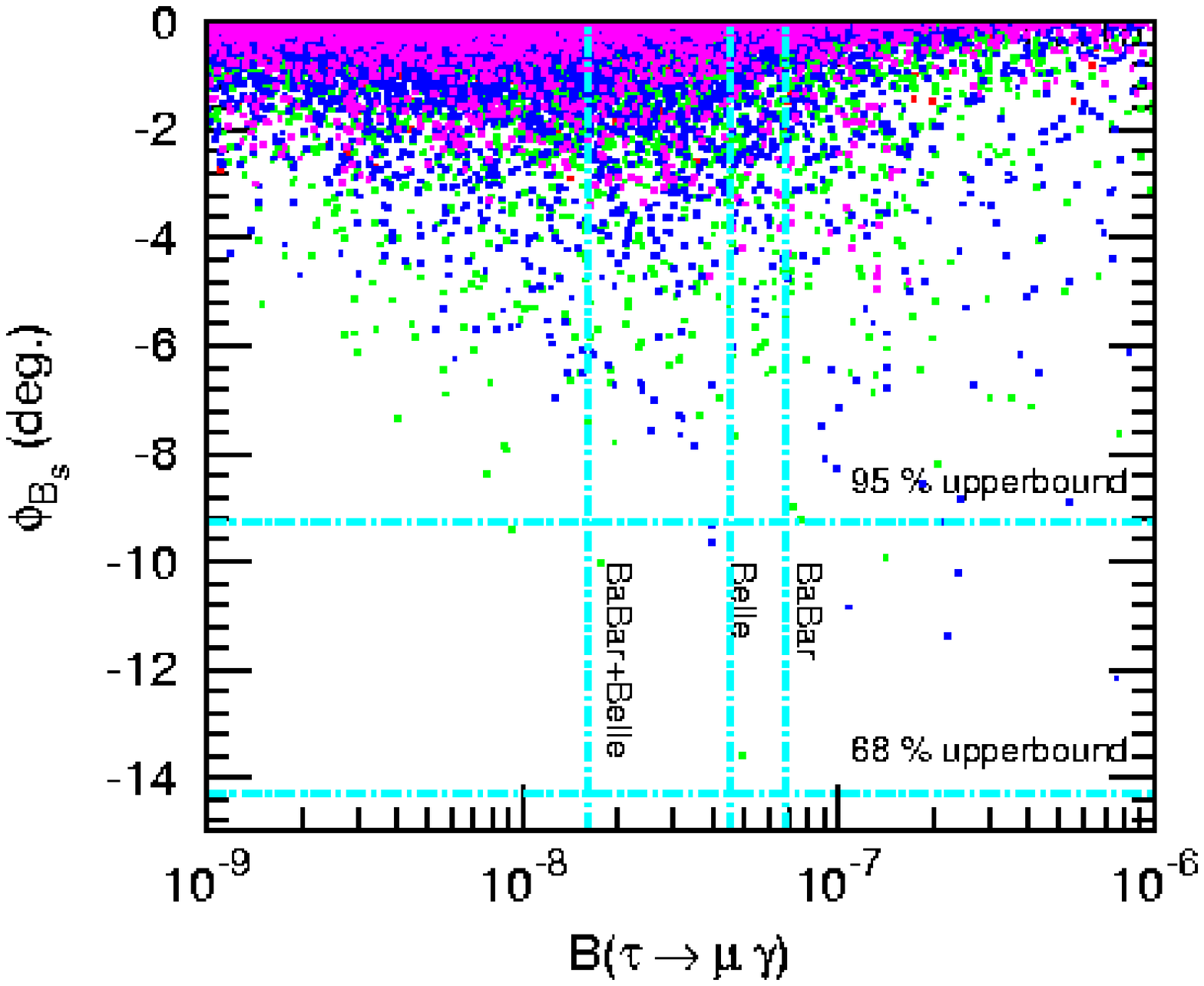}
\caption{\label{tmg-fbs-mn} 
Scatter plots on plane of $\phi_{B_s}$
and $Br(\tau\rightarrow\mu\gamma)$ for  $M_{N_3}=2$(red), $4$(green), $6$(blue), $8$(purple)$\times 10^{14}$~GeV. Here, we imposed the Higgs boson mass bound. The SUSY breaking parameters are given from Eq.~(\ref{region2}). 
}
\end{center}
\end{figure}

In the above discussion we took $M_{N_3}=6\times 10^{14}$~GeV for the
right-handed tau neutrino mass. When $M_{N_3}$ is larger, the neutrino
Yukawa coupling is larger with the left-handed tau neutrino mass
fixed. However, the scalar mass squareds for the ${\bf
  5^\star}$-dimensional multiplets may become negative above the GUT
scale. In the case, the flavor-violating mass terms of the ${\bf
  5^\star}$-dimensional multiplets rather become smaller so that the
deviation of $\phi_s$ is not enhanced.  In fact, we found it difficult
to get larger deviation of $\phi_{B_s}$ by raising $M_{N_3}$.  In
Fig.~\ref{tmg-fbs-mn}, $\phi_{B_s}$ and $Br(\tau\rightarrow\mu\gamma)$
are shown for several values of $M_{N_3}$. The input parameters for
the SUSY breaking parameters are taken here as
\begin{eqnarray}
0<m_0,~ M_{1/2}<3~{\rm TeV},\nonumber\\
-3<A_0/m_0<+3,\nonumber\\
5<\tan\beta<40\ ,
\label{region2}
\end{eqnarray}
which are wider than in Eq.~(\ref{region}).

In summary, we discussed how large correction in $\phi_{B_s}$ is
possible in the SUSY SU(5) GUT with right-handed neutrinos. Here, we
assumed the supergravity-like boundary condition for the SUSY-breaking
terms. We found that the $95\%$ probability region derived by the {\bf
  Ut}{\it fit} collaboration is marginal in this model. 

\section*{Acknowledgement} This work was supported by World Premier
International Center Initiative (WPI Program), MEXT, Japan.  The work
of JH was also supported in part by the Grant-in-Aid for Science
Research, Japan Society for the Promotion of Science (No.~20244037 and
No.~2054252).

\end{document}